\documentclass [a4paper]{article}
\usepackage{epsfig}

\usepackage{epstopdf}

\begin{document}
\noindent

\setlength{\textwidth}{6 in}
\setlength{\textheight}{10 in}
\setlength{\topmargin}{-0.5in}

\title{The Quantum Zeno Effect - Watched Pots in the Quantum World.}

\author{Anu Venugopalan}
\maketitle


\begin{abstract} 

In the 5th century B.C.,the philosopher and logician Zeno of Elea posed several paradoxes which remained unresolved for over two thousand five hundred years. The $20^{th}$ century saw some resolutions to Zeno's mind boggling problems. This long journey saw many significant milestones in the form of discoveries like the tools of converging series and theories on infinite sets in mathematics. In recent times, the  Zeno effect made an intriguing appearance in a rather unlikely place - a situation involving the time evolution of a quantum system, which is subject to ``observations'' over a period of time. Leonid Khalfin working in the former USSR in the 1960s and ECG Sudarshan and B. Misra at the University of Texas, Austin, first drew attention to this problem. In 1977, ECG Sudarshan and B. Misra published a paper on the quantum Zeno effect, called ``The Zeno's paradox in quantum theory''. Their fascinating result revealed the bizarre workings of the quantum world. Misra and Sudarshan's 1977 paper activated over two decades of theoretical and experimental explorations into the subject and still continues to evoke a lot of interest. In the following, the quantum Zeno effect is described and a brief outline of some of the work following Misra and Sudarshan's paper is given. The quantum Zeno effect is yet another example of the myriad unimaginable possibilities that lie waiting in the magical world of the quantum.

\end{abstract}

\begin{flushleft}
{\bf Keywords} 
\end{flushleft}
{\bf Zeno paradox, unstable systems, survival probability, quantum evolution, quantum measurements, continuous measurements, quantum Zeno effect, quantum anti-Zeno effect.}

\section{Introduction}
The quantum Zeno effect gets its name from the Greek philosopher Zeno of Elea. Zeno lived in the 5th century B. C. in the Greek colony of Elea (now in southern Italy). Zeno was known to be the most brilliant disciple of Parmenides, a very prominent figure of the Eleatic School of philosophers. Aristotle is supposed to have credited Zeno with having invented the method of the ``dialectic'', where two speakers alternately attack and defend a thesis. Zeno is also credited with  inventing the argument form 'reductio ad absurdum'. However, Zeno is most famous for his four paradoxes of motion which he developed as arguments against the then existing notions of space and time. A description of the quantum Zeno effect does not necessarily require a prior knowledge of the classical paradoxes of Zeno. However, as an interesting historical reference, we will briefly describe particular versions of the four classical paradoxes before embarking on a description of the quantum Zeno effect itself. The four classical paradoxes of Zeno  popularly go by the names: (i) Achilles and the tortoise, (ii) The arrow, (iii) the dichotomy, and (iv) the sophisticated stadium. In the following, we will briefly describe them, one by one.
\subsection{Achilles and the Tortoise}
Achilles has to race a tortoise. Since Achilles is obviously faster than his opponent, the tortoise is given a head start. Let us suppose that the race begins at t=0, when Achilles is located at $x=0$, and the tortoise, with its head start stands at $x=x_{0}$ [see Fig.1(a)]. Zeno argued that the running Achilles could never catch up with the tortoise because he must first reach where the tortoise started. For instance, when Achilles reaches $x_{0}$ at time $t_{0}$, the tortoise has already progressed to $x_{1}$ ($>x_{0}$). By the time Achilles reaches $x_{1}$, at time $t_{1}$, the tortoise has crawled up to $x_{2}$. Thus, whenever Achilles reaches $x_{i}$, the tortoise would have moved to  $x_{i+1}$, and Achilles would need time $t_{i+1}-t_{i}$ before he gets to that point (by which time the tortoise would have  moved on ahead!). Zeno argues that for Achilles to reach the tortoise (or to overtake it), he must perform an infinite sum of such time increments, $t_{i+1}-t_{i}$, or spatial increments,  $x_{i+1}-x_{i}$ for all $i$ upto $\infty$.
If space and time are considered to be continuous, these increments would tend to zero duration, and if they are considered discrete, the increment would be of finite duration. (Note that in the continuous case, the increments {\it tend to} zero but will not be actually zero for any finite $i$). Zeno argues that if space and time are continuous, an infinite sum of elements tending towards zero length (duration) must have a total of zero length (duration). Alternatively, if space and time are discrete, then an infinite sum of finite elements must be of infinite length (duration). Since Achilles is ``seen'' to overtake the tortoise, the above arguments fail. Thus, the seemingly absurd  conclusion that follows is that both space and time can neither be continuous nor discrete. This compels us to consider the notion that space and time are illusory- hence, so is everything we see!
\subsection{ The Arrow}
 Imagine an arrow flying through space [see Fig.1(b)]. Time is considered to be made up of "instants". These instants are defined as the smallest measure and they are indivisible. At any instant of time, when the arrow is observed, it must be seen to occupy a space equal to itself. If it is ``seen'' to ``move'' at any instant, it means that the observer can divide an instant into a time when the arrow was ``here'' and a time when the arrow was ``there''. This would mean that the instant of time consists of parts, which violates its basic definition (that of it being indivisible). Thus, Zeno asserts that there are no instants of time when the arrow does move. This, again, leads to the conclusion that  the arrow is always at rest and that all motion is illusory.

\subsection{The Dichotomy}
 Imagine that Achilles wants to run from point {\bf A} to point {\bf B}[see Fig.1(c)]. Before he can cover half the distance to the end, he must cover the first quarter. Before this, he must cover the first eighth, and before that the first sixteenth, and so on. Before Achilles can cover any distance at all he must cover an infinite number of smaller spatial separations. Zeno argued that covering these infinite numbers of small spatial separations in a finite time would be impossible. Thus, it can be concluded that Achilles can never get started. But Achilles is ``seen'' to move. Thus, once again, the argument compels us to conclude that all motion is an illusion.This paradoxical argument is called the Dichotomy because it involves repeatedly splitting a distance into two parts. It contains some of the same elements as the Achilles and the Tortoise paradox, but with a more apparent conclusion of motionlessness.

\subsection{ The Sophisticated stadium}
This is the last of Zeno's four paradoxes. Let us suppose that space and time are discrete in nature and that motion consists in occupying different spatial points at different times. For simplicity, consider objects moving at a minimal speed,  $v$, of one fundamental spatial distance per fundamental temporal duration. Consider nine persons moving roughly collinearly [See Fig 1(d)]. Persons $a_i$ are all stationary, while persons $b_i$ move past them to the left, with the velocity, $-v$. At the same time, persons $c_i$ move to the right past the $a_i$s with velocity $v$. Say, at time $t=0$ the configuration is as shown in Fig.1(d (i)). One fundamental unit of time later, the configuration in Fig $1(d(ii))$ is achieved. One can see that $c_3$ has passed by $b_2$, but there was never an instance when $c_3$ was lined up with $b_2$. Thus, it can be argued that there is no time at which the actual passing occurs, and hence, it never happened! Another way to bring attention to this paradoxical situation is to note that in Fig. 1 d(ii), each {\bf B} has passed twice as many {\bf Cs} as {\bf  As}. Thus, one might conclude that it takes row {\bf B} twice as long to pass row {\bf A} as it does to pass row {\bf C}. However, the time for rows {\bf B} and {\bf C} to reach the position of row A is the same. So, it appears that half the time is equal to twice the time!

\bigskip
\noindent
The four paradoxes, summarized above in their barest forms, foxed and confused mathematicians and philosophers for over two millenia. It was only after the development of the calculus of infinitesimals by Leibniz and Newton, the concept of functions, limits, continuity, infinite series, and convergence, that a satisfactory resolution to Zeno's paradoxes came about. However, even today, inspite of our familiarity with these modern  mathematical ideas and concepts, there is a continuing debate about the validity of Zeno's paradoxes and their various resolutions. We end our brief introduction to the ``classical'' paradoxes of Zeno at this point and will describe the {\em Quantum Zeno effect} - the main subject of this article, in the next section.

\section{The Quantum Zeno Effect}

As mentioned in the previous section, the Zeno paradox in quantum systems was brought into focus in the 1960s by  Leonid A. Khalfin, working in the former USSR, and by E.C.G. Sudarshan and Baidyanath Misra, working in the USA during the 1970s. Misra and Sudarshan's paper entitled ``The Zeno's  paradox in Quantum Theory'' in the Journal of Mathematical Physics first introduced the name "Zeno's paradox" for the effect studied. "Quantum Zeno Effect" (QZE) is a term more commonly used these days to describe similar situations in various quantum systems. In order to understand the essence of the QZE, it is necessary to first briefly review some basic concepts of quantum mechanics and quantum measurement.

\subsection{Quantum Measurement}

Quantum mechanics is currently accepted as the most elegant and satisfying description of phenomena at the atomic scale. Since our larger, familiar ``macro'' world is eventually composed of elements of the ``micro'' world which is described by quantum mechanics, quantum theory is also, inevitably, the fundamental theory of nature. Though stunningly powerful, the quantum mechanical view of the world has compelled us to re shape and revise our ideas of reality and notions of cause, effect and measurement. Without going into too many details of the various conceptual difficulties of quantum mechanics, we will only focus on the quantum mechanical description of {\em measurement} which is of direct relevance to the description of the Quantum Zeno effect.

The quantum mechanical description of a system is contained in its wavefunction or state vector, $|\psi \rangle$, which lives in a abstract ``Hilbert space''.  The dynamics of the wavefunction is governed by the Schr\"{o}dinger equation:
\begin{equation}
i\hbar \frac{d}{dt}|\psi\rangle=H|\psi\rangle,
\end{equation}
where $H$ is the Hamiltonian operator, and the equation is linear, deterministic and the time evolution governed by it is {\em unitary}. Unitary evolutions preserve probabilities. Examples of unitary transformations are rotations and reflections. Dynamical variables or {\em observables} are represented in quantum mechanics by {\em linear Hermitian operators}, which act on the state vector. An operator, $\hat{A}$, corresponding to a dynamical quantity, $A$, is associated with {\em eigenvalues} $ a_{i}$s and corresponding {\em eigenvectors}, $\{ | \alpha_{i} \rangle \}$, which form a {\em complete orthonormal set}. Any arbitrary state vector, $|\psi \rangle$, in general, can be represented by a linear superposition of these eigenvectors, or, for that matter, a  combination of any orthonormal set of basis vectors in Hilbert space. Thus, one can write $|\psi\rangle=\Sigma c_{i}|\alpha_{i}\rangle$. A basic postulate of quantum mechanics regarding {\em measurement} is that any measurement of the quantity A can only yield {\em one} of the eigenvalues, $a_{i}$s,  but the result is not definite in the sense that different measurements for the quantum state  $|\psi\rangle$ can yield different eigenvalues. Quantum theory predicts only that the {\em probability} of obtaining eigenvalue $a_{i}$ is $|c_{i}|^{2}$. Quantum theory defines the {\em expectation value} of the operator $\hat{A}$ as:  
\begin{equation}
\langle\hat{A}\rangle=\langle\psi|\hat{A}|\psi\rangle=\Sigma{ a_{i}|c_{i}|^{2}}.
\end{equation}
In terms of the density matrix $\hat{\rho}=|\psi\rangle\langle\psi|$, an equivalent formula for the expectation value is:
\begin{equation}
\langle\hat{A}\rangle=Trace \{\hat{A} \hat{\rho} \}.
\end{equation}

An additional postulate of quantum mechanics is that the measurement of an observable A, which yields one of the eigenvalues $a_{i}$ ( with probability $|c_{i}|^{2}$) culminates with the {\em reduction } or {\em collapse } of the state vector $|\psi \rangle$ to the eigenstate $| \alpha_{i}\rangle$. This means that every term in the linear superposition vanishes, except one. This reduction is a {\em non unitary process} and hence in complete contrast to the unitary dynamics of quantum mechanics predicted by the Schr\"{o}dinger equation and this is where the crux of the conceptual difficulties encountered in quantum theory lies. For now we just accept this as a basic  postulate of quantum theory (also called the {\em projection postulate}) and go on to describe the quantum Zeno effect.

\subsection{The Result of Misra and Sudarshan}

The quantum Zeno effect (or paradox) was the name given by Misra and Sudarshan to the phenomenon of the {\em inhibition} of transitions between quantum states by frequent {\em measurements}. For their study, they considered the decay of an unstable state, such as an unstable particle, like a radioactive nucleus. The classic model of any system decay is an exponential function of time in most situations. We are all familiar with the radioactive decay law
\begin{equation}
N(t)=N(t_{0})e^{-\lambda(t-t_{0})},
\end{equation}
where $N(t)$ is the number of nuclei that have not decayed after time $t$, and $\lambda$ is a constant which depends on the properties of the species of nuclei. While the decay of a quantum system is similar to this classic model of exponential decay, there have been theoretical studies that show that in certain timescales (specifically, for {\em very  short} and {\em very long} times as measured from the instant of preparation of the state of the system), there can be a deviation from the familiar exponential decay law. Infact, recently this theoretical deviation from the exponential decay law has also been confirmed experimentally in quantum tunnelling experiments with ultra cold atoms by a group at the University of Texas, USA. It is in these special time regimes that we see manifestations of the quantum Zeno effect. Consider the decay of an unstable quantum state. Let $\psi_{0}$ be the (undecayed) state of the system at $t=0$ and $\psi(t)$ be the state at any later time $t$. The evolution of the state is governed by a unitary operator, $U(t)$, where
\begin{equation}
 U(t)=e^{-iHt},
\end{equation}
(here $\hbar=1$) and 
\begin{equation}
|\psi(t)\rangle=U(t)|\psi_{0}\rangle,
\end{equation}
 $H$ being the Hamiltonian of the system.  As discussed in the previous section, any observation that the state has not decayed will cause a collapse (reduction) of the wavefunction to the undecayed state. The {\em survival probability}, $P(t)$, i.e., the probability that the system is still in the undecayed state,  will be the modulus squared of the {\em survival amplitude} and can be written as:
\begin{equation}
P(t)= |\langle\psi_{0}|U(t)|\psi_{0}\rangle|^{2}.
\end{equation}
Now, (7) can be expanded as:
\begin{equation}
P(t)=1-t^{2}\big(\langle\psi_{0}|H^{2}|\psi_{0}\rangle-\langle\psi_{0}|H|\psi_{0}\rangle^{2}\big) + .....
\end{equation}
If 
\begin{equation}
\Delta H=\sqrt{\langle\psi_{0}|H^{2}|\psi_{0}\rangle-\langle\psi_{0}|H|\psi_{0}\rangle^{2}},
\end{equation}
then the survival probability in the {\em short time limit} can be rewritten as:
\begin{equation}
P(t) \approx 1-t^{2} (\Delta H)^{2} + ......
\end{equation}
If we define $\tau_{Z}=1/\Delta H$ as the {\em Zeno time}, this gives:
\begin{equation}
P(t) \approx 1-\frac{t^{2}}{\tau_{Z}^{2}} + ...
\end{equation}
which, for short times, can be written as:
\begin{equation}
P(t) \approx \big(1-\frac{t^{2}}{\tau_{Z}^{2}}\big).
\end{equation}
The above expression shows that the short-time quantum decay is not exponential in time, but {\em quadratic}. Now, let us suppose that one makes $N$ equally spaced {\em measurements} over the time period $[0,T]$. If $\tau$ is the time interval between two measurements, then $T=N\tau$. Let us assume that the measurements are made at times $T/N$, $2T/N$, $3T/N$, .....$(N-1)T/N$ and $T$ and are {\em instantaneous}. So, essentially this describes an alternate sequence of unitary evolutions (each lasting time $\tau$) followed by a collapse (the basic postulate of quantum measurement). The survival probability after $N$ measurements, or after time $T$ can be written as:
\begin{equation}
P^{N}(T)=[P(\tau)]^{N}=\big(1-\frac{T^{2}}{N^{2}\tau_{Z}^{2}} \big)^{N}.
\end{equation}
It can be seen that in the limit of {\em continuous measurements}, i.e., when $N \rightarrow \infty$, 
 \begin{equation}
  \lim_{N \rightarrow \infty} P^{N}(T)= 1.
\end{equation}
Thus {\em the probability that the state will survive for a time $T$ goes to ${\bf 1}$ in the limit $N \rightarrow \infty$}. This means that continuous measurements actually prevent the system from ever decaying! So, much like the motionless arrow in Zeno's paradox, the system never decays, or a ``watched pot never boils''. Now, can we see this happening in a real experiment? Unfortunately, in  spontaneous decay this effect is very difficult to observe for reasons that we will briefly discuss in a later section. No experiment has been able to probe this regime to observe the inhibition of the decay of an {\em unstable particle} like a radioactive nucleus, as yet. However, as mentioned earlier, recently there have been experimental groups which have reported the observation of the quantum Zeno effect in  unstable systems comprising of trapped ultra cold atoms which undergo quantum mechanical {\em tunnelling}. We will discuss these later. First, we will describe an  earlier experiment carried out by Itano et al in 1990 at the experimental group headed by Wineland at the National Institute of Standards and Technology, Boulder, Colorado. This was the first instance when a manifestation of the quantum Zeno effect was successfully demonstrated experimentally. This is discussed in the next section.

\subsection{Experimental manifestations of the Quantum Zeno Effect}
\subsubsection{The Experiment of Itano et al.}

Following an original proposal by Cook, Itano et al, at the National Institute of Standards and Technology, Boulder, Colorado, experimentally demonstrated the occurrence of the Quantum Zeno effect in {\em induced} transitions between two quantum states. Unlike the case studied by Misra and Sudarshan, this is a situation where there is no spontaneous decay of an unstable system  but an {\em induced} transition between two states of a system. The experiment of Itano et al tested the inhibition of the induced radio frequency transition between two hyperfine levels of a Berylium ion, caused by {\em frequent measurements} of the level population using optical pulses. The experiment can be understood as follows [see Fig.2]: Consider a two-level system, with the levels labelled as $1$ and $2$. Assume that the system can be driven from level $1$ to level $2$ by applying a {\em resonant radio frequency pulse}. Assume that it is possible to make instantaneous measurements of the state of the system, i.e., to ascertain  whether the system is in level $1$ or in level $2$. In order to observe the level populations, level 1 is connected by an optical transition to an additional level $3$ such that level $3$ can decay only to level $1$. Spontaneous decay  from level $2$ to level $1$ is negligible. The measurements are carried out (during the evolution under the resonant radio frequency pulse) by driving the $1 \rightarrow 3$ transition with $N$ equispaced short optical pulses and observing the presence (or absence) of spontaneously emitted photons from level $3$ to level $1$. Such a situation was created in a real experimental system with a trapped Berylium ion where  appropriate energy levels of the ion could be chosen to correspond to the $1, 2$ and $3$ levels described above. In recent times, trapped ions and atoms have become very popular systems for carrying out many experiments that test fundamental issues in quantum mechanics. They are considered ``clean'' systems that can be observed for long periods of time and isolated from noise. Moreover, their  energy levels can be easily manipulated with appropriate radio frequency and optical pulses. 

Suppose the ion is in level $1$ at a time $t=0$. An rf field having resonance frequency $\Omega=(E_{2}-E_{1})/\hbar)$ is applied to the system and this creates a state which is a {\em coherent superposition} of states $1$ and $2$. The dynamics of a two-level system in the presence of resonant driving field is well-studied and understood. The frequency $\Omega$ is called the Rabi frequency. An on-resonance ``$\pi$ $pulse$'' is a pulse of duration $T=\pi/\Omega$ and takes the ion from level $1$ to level $2$. If $P_{2}(t)$ is the probability at time $t$ for the ion to be at level $2$, then $P_{2}(T)=1$. This would be the situation when no ``measurement pulses'' are applied. In the experiment, $N$ measurement pulses are applied (which connect level $1$ to level $3$ through an optical pulse each time), {\em within} time $T$, i.e.,  at times $\tau_{k}=kT/N, k=1,2,3,....N$. Note that the dynamics of the two-level system driven by the resonant rf field is unitary and can be described quantum mechanically using the Bloch vector representation. The nonunitary {\em projection postulate} (or the {\em collapse} induced by quantum measurement) is incorporated each time a measurement is made. It is easy to show that at the end of $N$ measurements, i.e., at the end of the rf pulse at time $T$, the probability $P_{2}(T)$, which  corresponds to the {\em population} of level $2$ is given by:

\begin{equation}
P_{2}(T)=\frac{1}{2}[1-\cos^{N}(\pi/N)].
\end{equation}
For large N, i.e., in the limit of {\em continuous measurements}, one can see that
\begin{equation}
P_{2}(T)=\lim_{N \rightarrow \infty}\frac{1}{2}[1-\cos^{N}(\pi/N) \approx 0.
\end{equation}
Clearly, the continuous measurements described above {\em inhibit} the induced transition from $1$ to $2$, making the system ``freeze'' in level $1$. This  effect showed itself up in the real experimental observations  of Itano et al. Thus, although it was not seen in the decay of a unstable particle, the experiment of Itano et al was the first real demonstration of the quantum Zeno effect. Interestingly, the experiment was followed by a slew of papers where many issues were raised regarding the actual dynamics of the mechanism explored in the experiment by Itano et al. The experiment was critically analyzed from the point of view of {\em quantum measurements} and questions were raised regarding whether or not the {\em collapse} postulate plays any role at all in the outcome of the experiment. Many physicists assert that the quantum Zeno effect is simply a consequence of the unitary dynamics of conventional quantum mechanics and need not involve the non unitary collapse of quantum measurement. However, since the projection postulate of conventional quantum measurement theory also successfully described the outcome of this experiment, it is valid to see the experiment of Itano et al as a demonstration of the inhibition of transition due to frequent measurements, or the quantum Zeno effect. In the next section we describe one more experimental manifestation of the QZE.

\subsubsection{The Experiment of Kwiat et al}
In 1995, Paul Kwiat and his group at the University of Innsbruck realized a version of the Quantum Zeno Effect in the laboratory using the polarization directions of single photon states. Their experiment was based on an example first suggested by Asher Peres in 1980. Consider plane polarized light. It can have two possible polarization directions, say, "vertical" and "horizontal". We know that when such a beam passes through an optically active liquid (e.g., sugar solution)
its plane of polarization is rotated by a small angle (which depends on the concentration of the sugar solution, for example). Consider a photon directed through a series of such "rotators" so that each slightly rotates its polarization direction so that an initially vertically polarized photon ends up horizontally polarized. At the end of this series of rotators, the photon encounters a polarizer. A polarizer is a device that transmits photons with one kind of polarization but absorbs photons with the perpendicular polarization. An ideal Nicol prism acts as a polarizer (or analyzer). Now let us suppose that an experiment is set up with six rotators, each of which rotates the plane of polarization of a vertically polarized photon by $15^{0}$. At the end of this series is a polarizer which transmits only vertically  polarized light, which is then detected by a photon detector [See Fig 3]. It is obvious that in the above set up the photon will never get to the detector as its polarization will have turned by $90^{0}$ after passing through the six rotators and become horizontal. To implement the Zeno effect, Paul Kwiat and his colleagues sought to {\em inhibit} this step wise rotation of the polarization, or the evolution of the polarization state from the vertical to the horizontal, by {\em measurements} of the polarization state.
Kwiat et al realized this by interspersing a vertical polarizer between each rotator [see Fig.3]. If the first rotator rotates the plane of polarization by an angle $\alpha$, then the vertical polarizer kept after it will transmit the photon with a probability $\cos^{2}{\alpha}$, and the original vertical polarization would have been restored (this can be recognized as the well known Malus' law). At the second rotator the polarization is once again turned by $\alpha$ and it then encounters  the second polarizer where it will be tramsmitted with a probability $\cos^{2}{\alpha}$ and the vertical polarization will be, once again, restored. This process repeats till the photon comes to the final polarizer. If $\alpha=15^{0},(\cos^{2}{\alpha})^{6}=\frac{2}{3}$. Thus an incident photon has $\frac{2}{3}^{rd}$  chance of being transmitted through all six inserted polarizers and making it to the detector. It can be easily seen that if one increases the number of stages, decreasing the rotation angle at each stage, the probability of transmitting the photon to the detector increases. If there were an infinite number of stages, the photon would always get through and hence the rotation of the plane of polarization would be completely inhibited - the Zeno effect! In the actual experiment, Kwiat et al created single photon states using a nonlinear crystal. Thus, like the experiment of Itano et al, the experiment of Kwiat et al demonstrated the suppression of evolution in a driven two state system through frequent measurements. What about the QZE in unstable systems?

\subsection{The Result of Kurizki and Koffman - The Anti-Zeno Effect}
In the previous section we have seen experimental evidences of the quantum Zeno effect in induced transitions between two quantum states.  A natural question that arises, then, is whether the Zeno effect can be used (in a real experiment) some day to "freeze" radioactive nuclear decay. For the past three decades it seemed that the answer to the question was a 'yes', provided one had the experimental technology and sophistication to perform successive, "frequent enough" measurements. However, recent  work by Gershon Kurizki and Abraham Koffman at the Weizmann Institute of Science, Israel, has shown that such a freezing may actually not be possible at all. Kurizki and Koffman have argued that there is an "{\em Anti-Zeno Effect}" which infact {\em enhances} the decay of unstable particles instead of inhibiting it! According to their calculations, the ability to "freeze" the evolution through frequent measurements depends on the ratio between the "memory time" of the decay, and the time interval between successive measurements. Every decay process has a "memory time". This memory time is the time following a quantum event in which the particle can still return to its initial state. In the case of radioactive decay, for instance, the memory time is the period in which the radiation has not yet escaped from the atom, allowing the system to "remember" its state prior to the decay. Typically, this memory time for an unstable particle is less than one billionth of a billionth of a second. Kurizki and Koffman argue that frequent measurements in this time scale (if it were possible)  would cause more particles to be created. This would interfere with, and essentially destroy the original system, making it meaningless to ask whether the decay has frozen or not. On the other hand, if the time interval between measurements is longer than the memory time (i.e., observations are not fast enough for the "expected" QZE), the rate of decay {\em increases} and one would have the {\em Anti Zeno effect}.
While we will not go into the details of their work, we can state that {\em the surprising conclusion of the research of Kurizki and Koffmann is that the Anti-Zeno effect (i.e., the increase of decay through frequent measurement) can occur in all processes of decay, while the Zeno effect which would slow down and even stop decay requires conditions that are much rarer.}

While the predictions of Kurizki and Koffman are yet to be experimentally verified on an unstable system like a radioactive nucleus, recent experiments by Mark Raizen and his colleagues at the University of Texas, Austin have demonstrated the quantum Zeno and the quantum Anti Zeno effects in the tunnelling behaviour of cold trapped atoms. Raizen's team trapped sodium atoms in a "light wave". Such a system, if left alone, will slowly decay as individual atoms escape through quantum mechanical tunnelling through an energy barrier which would be classically insurmountable. Through ingenious experimental techniques, the team showed that the tunnelling rate slowed down dramatically when they 'measured' the system every millionth of a second - the quantum Zeno effect! When they measured the system every five millionth of a second, the tunnelling rate increased - the quantum Anti Zeno effect! It is interesting to notice the happy coincidence that this spectacular experimental test of the QZE which is closest in spirit to the original proposal of Misra and Sudarshan was performed at the University of Texas, Austin - the very same place from where Misra and Sudarshan published their work, almost three decades ago.
\section{Conclusions}

Much water has flown under the bridge since Zeno wondered about Achilles and the  tortoise at the dawn of civilization. Though mathematicians  have solved the classical  paradoxes of Zeno long ago by introducing the concept of real numbers, limits, continuity and calculus to describe quantities of duration and distance, the notion of freezing motion by continuous observation has  turned out to be a very real effect in the strange world of quantum physics. Spectacular experiments bear testimony to the  reality of this effect in  the quantum domain and the fields  of atom optics and cold trapped ions continue to spring up tantalizing new surprises every day. While in the classical world Achilles overtakes the tortoise and all is well with the world, in the mysterious land of the quantum, watched pots stop boiling (or boil faster, maybe!) and the ghost of Zeno continues to make its presence felt in unimaginably interesting  ways.

\begin{flushleft}
{\bf The author would like to thank Ragothaman Yennamalli and Vivek for help with the figures.}
\end{flushleft}
\begin{flushleft}
{\bf Suggested Reading}
\end{flushleft}
\begin{flushleft}
The topics touched upon in this article cover several  references. The interested reader may look at some of the following:
\end{flushleft}
\begin{flushleft}
{\bf [1]} {\em A watched atom never decays}, {\bf New Scientist 10th March 1990}\\
\vspace{0.1in}
\noindent
{\bf [2]} {\em Accelerated decay}, {\bf New Scientist, 3rd June 2000}\\
\vspace{0.1in}
\noindent
{\bf [3]} {\em Zeno paradox in Quantum Theory}, {\bf Asher Peres, American Journal of Physics (1980),  p 931}\\
\vspace{0.1in}
\noindent
{\bf [4]} {\em Quantum Zeno Effect}, {\bf ``Decoherence and  Quantum Measurements"}, {\bf Namiki, Pascazio and Nakazato, World Scientific (1997)}, Chapter 8 \\
\vspace{0.1in}
\noindent
{\bf [5]} The description of the classical paradoxes of Zeno is based on information at the online site
 {\bf http://pirate.shu.edu/~wachsmut/ira/history/zeno.html} and {\bf Chapter 9} of the PhD thesis of  {\bf M.J Gagen}, entitled {\bf "Quantum measurement theory and the quantum Zeno effect" (PhD thesis, University of  Queensland, Australia 1993)} \\
\vspace{0.1in}
\noindent
{\bf [6]} The section on the experiment of Kwiat et al is based on material posted at 
{\bf http://www.fortunecity.com/emachines/e11/86/seedark.html} entitled {\bf "Quantum Seeing in the Dark", by Paul Kwiat, Harald Weinfurter and Anton Zeilinger}
\end{flushleft}
\pagebreak
\begin{figure}
\resizebox {5 in}{3.5 in}{\includegraphics[width=0.9\textwidth]{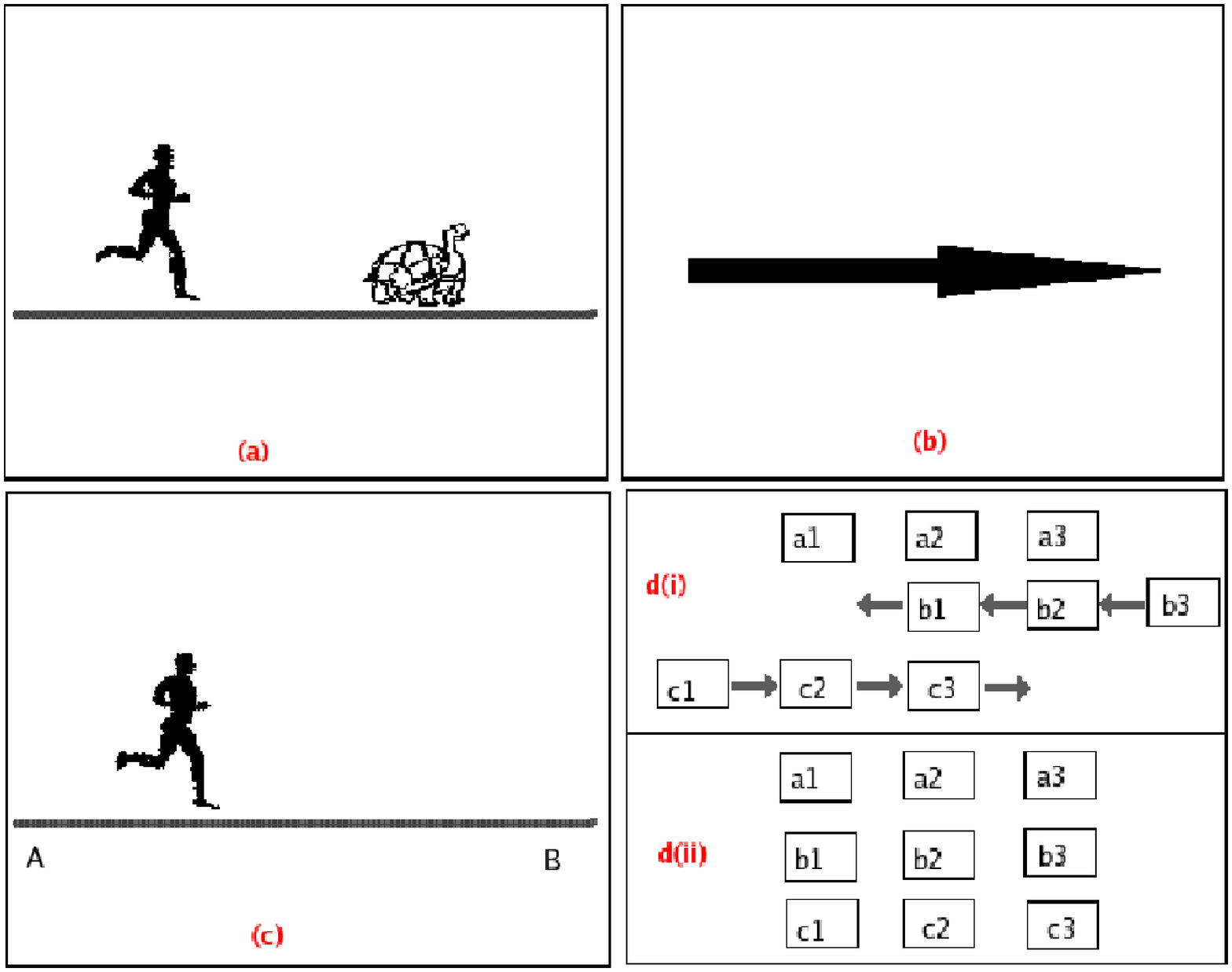}}
\caption{\bf (a) Achilles and the Tortoise, (b) The Arrow, (c) The Dichotomy, (d) The Sophisticated Stadium}
\vspace{1 in}
\resizebox {5 in}{3.5 in}{\includegraphics[width=0.9\textwidth]{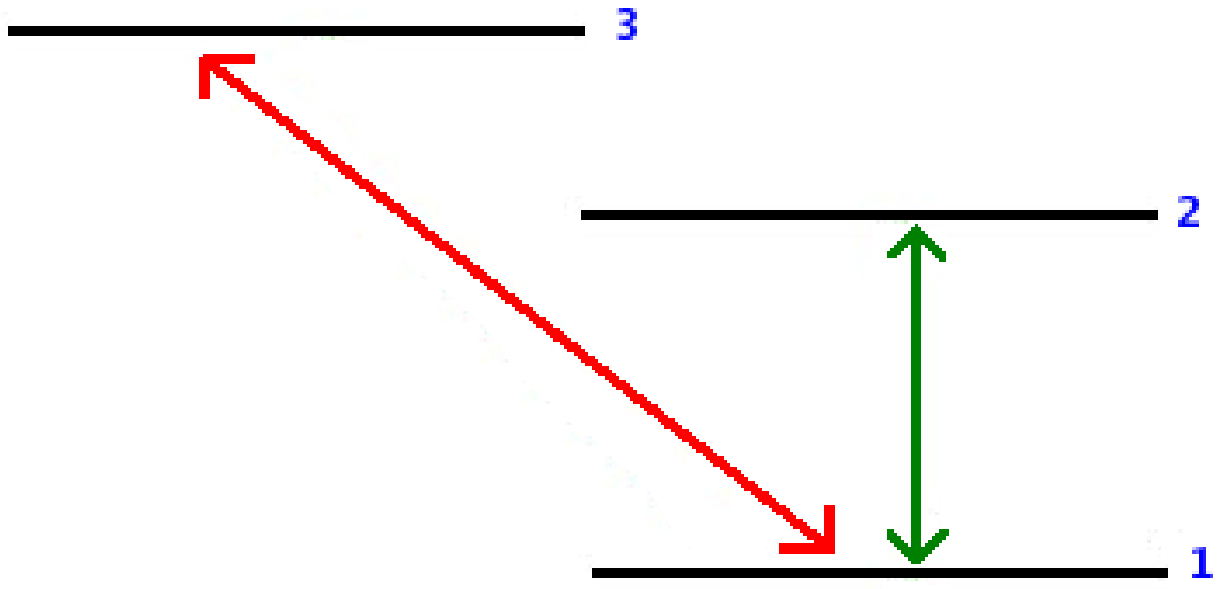}}
\caption{\bf Quantum Zeno Effect in Induced Transitions between energy levels - the experimental system of Itano et al}
\hfill
\end{figure}

\pagebreak
\begin{figure}
\resizebox {5 in}{1.5 in}{\includegraphics[width=1.0\textwidth]{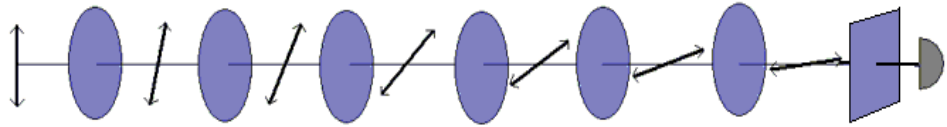}}
\caption{\bf Six rotators turn the polarisation by 15 degrees at each stage such that a vertically polarized  photon changes to a horizontally polarised one.}
\hfill
\end{figure}

\vspace{-4 in}

\begin{figure}
\resizebox {5 in}{1.5 in}{\includegraphics[width=1.0\textwidth]{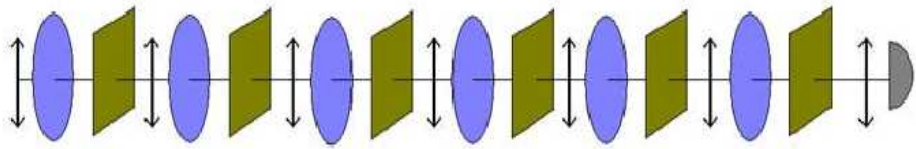}}
\caption{\bf Quantum Zeno Effect:Interspersing a polariser after each rotator inhibits the polarisation state from changing.}
\hfill
\end{figure}

\end{document}